\documentclass[10pt,journal,letterpaper]{IEEEtran}
\usepackage[utf8]{inputenc}

\usepackage{graphicx}	
\usepackage{adjustbox}
\usepackage{array}
\usepackage{amssymb}
\usepackage{pifont}
\usepackage{bm}
\usepackage{amsmath}
\usepackage{amsthm}
\usepackage{algorithm,algorithmic}
\usepackage[dvipsnames]{xcolor}
\usepackage{colortbl}
\usepackage{enumitem}   
\usepackage{tabulary}
\usepackage{booktabs}
\usepackage{float}
\usepackage{subcaption}
\usepackage{tikz}
\usepackage[]{cryptocode}
\usepackage{multirow}
\usepackage{arydshln}
\usepackage{makecell}
\usepackage{rotating}
\usepackage{tikz}
\usepackage[most]{tcolorbox}
\usepackage[dvipsnames]{xcolor}
\usepackage{tabularx}

\usepackage[hyphens]{url}
\usepackage[hidelinks]{hyperref}
\hypersetup{breaklinks=true}

\newcommand{\pie}[1]{%
\begin{tikzpicture}
 \draw (0,0) circle (0.6ex);\fill (0.6ex,0) arc (0:#1:0.6ex) -- (0,0) -- cycle;
\end{tikzpicture}%
}

\newcolumntype{R}[2]{%
    >{\adjustbox{angle=#1,lap=\width-(#2)}\bgroup}%
    l%
    <{\egroup}%
}
\newcommand*\rot{\multicolumn{1}{R{25}{1em}}}%
\newcommand{\etal}{\textit{et\,al.\;}}
\newcommand{\eg}{\textit{e.g.,\,}}

\newcommand{\ie}{\textit{i.e.,\,}}

\newcommand{\descr}[1]{\smallskip \noindent \textbf{#1}}

\newcommand{\descrit}[1]{\smallskip \noindent \textit{#1}}

\begin{document}
  \title{SoK: Privacy-Preserving Collaborative Tree-based Model Learning\thanks{To appear in the Proceedings on Privacy Enhancing Technologies (PoPETs),Vol. 2021, Issue 3}}

\author{
\IEEEauthorblockN{Sylvain Chatel\footnote{corresponding author}, Apostolos Pyrgelis, \rm Juan Ram\'{o}n Troncoso-Pastoriza, \rm Jean-Pierre Hubaux}

\IEEEauthorblockA{Laboratory for Data Security -- EPFL,\\first.last@epfl.ch}
}

\maketitle
\thispagestyle{plain}
\pagestyle{plain}

\begin{abstract}
Tree-based models are among the most efficient machine learning techniques for data mining nowadays due to their accuracy, interpretability, and simplicity. The recent orthogonal needs for more data and privacy protection call for collaborative privacy-preserving solutions. In this work, we survey the literature on distributed and privacy-preserving training of tree-based models and we systematize its knowledge based on four axes: the learning algorithm, the collaborative model, the protection mechanism, and the threat model. We use this to identify the strengths and limitations of these works and provide for the first time a framework analyzing the information leakage occurring in distributed tree-based model learning.
\end{abstract}

\section{Introduction}\label{sec:intro}
Tree-based models are currently among the most powerful data-mining methods. They are widely used in the industry\;\cite{nvidiaxgb,amazonxgb} and in machine-learning competitions\;\cite{Andriushchenko2019provably,sandulescu2016predicting}. 
These algorithms perform very well for tabular problems with numerical and categorical data, which places them in the top ten of machine-learning methods of 2017\;\cite{nugget2017} with numerous applications: fraud detection\;\cite{fang2020hybrid}, medical diagnosis\;\cite{alabdulkarim2019ppsdt}, and stock trading\;\cite{mabu2015ensemble}. 
An important feature of tree-based models is \emph{interpretability}, as it makes them an ideal candidate for the interpretable and explainable machine-learning quest of the last few decades\;\cite{molnar2020interpretable,doshi2017towards,freitas2014comprehensible}. Interpretability is related to the comprehensibility of a model, \ie to what extent the end-user is able to comprehend the model's learning rationale and to verify the soundness of its decisions. This is invaluable in several domains, \eg medicine and finance, where black-box machine-learning approaches are not acceptable. Indeed, when conducting medical experiments, researchers seek to identify the key factors that affect their outcome, not just the best predictive model on some data. As pointed out by Freitas\;\cite{freitas2014comprehensible}, the interpretability of decision trees makes them reliable, facilitates analysis, and orients future research by identifying points of interest.

Collaborative learning refers to the setting where a group of entities seeks to train a model on their joint data. Collaborative (also known as \emph{federated}\;\cite{mcmahan2017communication}) learning has received much traction, due to its applicability in settings where the data is scarce and distributed among multiple entities. For instance, in medical research, a rare disease is possibly not well represented in the patient data of one sole institution; hence the need for data sharing across diverse institutions to create a generalizable model with strong predictive performance.

However, collaborative learning raises privacy and security issues. The training data, \eg a hospital's patient medical records, is sensitive and cannot, without appropriate safeguards, be shared with other institutions. This is also reflected by the introduction of strict privacy regulations, such as HIPAA\;\cite{hipaa} and GDPR\;\cite{gdpr}, that forbid data sharing without proper anonymization or pseudonymization procedures. In particular, protecting data used in collaborative machine-learning pipelines is critical, as recent research introduces various successful privacy attacks\;\cite{hitaj2017deep,kairouz2019advances,melis2019exploiting,nasr2019comprehensive}. Any information exchanged while jointly training a machine-learning model can break the privacy of the training data or the resulting model. 

In this work, we perform a cross-field systematization of knowledge on privacy-preserving collaborative training of tree-based models such as decision-trees, random forests, and boosting. Our systematization is based on four axes: the learning algorithm, the collaborative model, the protection mechanism, and the threat model. Our study emphasizes the usage of privacy-enhancing technologies, showing their strengths and limitations.
We find that tensions arise as the learning, distributed environment, and privacy protections introduce new constraints. Elegant and efficient solutions exist but often at the cost of some information leakage, and the few end-to-end protected solutions are not amenable to all scenarios. Therefore, we also provide a framework that identifies the information leakage occurring during the collaborative training of tree-based models. Our systematization enables us to identify limitations such as relaxed threat models and the lack of end-to-end confidentiality, and overall highlights avenues for future work.

The remainder of the document is structured as follows. In \S\ref{sec:method}, we present an overview of our systematization methodology.
In \S\ref{sec:dt}, we provide background information on tree-based model learning. In \S\ref{sec:algo}, we expand on the learning algorithms and, in \S\ref{sec:collab}, on the types of collaborative settings. In \S\ref{sec:ppm}, we present the privacy-preserving mechanisms used in the literature and in \S\ref{sec:tm} the considered threat models. In \S\ref{sec:leak}, we present our leakage analysis framework. In \S\ref{sec:eval}, we give an overview of the evaluation conducted in the literature. Finally, in \S\ref{sec:open}, we discuss open challenges and we conclude in \S\ref{sec:conc}.

\subsection{Related Work}\label{sec:intro:rw}
There exist a few works similar to ours. 
While several works surveyed privacy-preserving data mining\;\cite{verykios2004state,ying2011state} or decision-trees\;\cite{truex2017privacy}, these works only sketched the collaborative and learning settings and did not delve into the challenges they induce. Recent works focused on federated learning\;\cite{li2019survey} and the security and privacy issues of machine learning\;\cite{papernot2018sok}, but none from the perspective of tree-based models. 
The literature on on decision-tree classification with differential privacy was investigated\;\cite{fletcher2019decision} but only in the centralized settings, where one entity holds the data (see \S\ref{sec:collab}). Similarly, a recent work systematized the knowledge on privacy-preserving decision-tree inference\;\cite{kiss2019sok}. Our work aims at bridging these gaps by systematizing the topic of privacy-preserving collaborative tree-based models focusing on the specific challenges induced by tree-learning in the distributed setting.

\subsection{Terminology}\label{sec:intro:not}
Let a \emph{party} be an entity owning a \emph{local} dataset consisting of samples with multiple features. A party seeks to participate in a tree-based model induction process with other parties. A \emph{miner} is an entity that performs computations. It either conducts the model training on the data owned by the parties or simply assists with intermediate computations. An \emph{aggregator} is an entity that combines, during the learning process, information from multiple parties. 
We also employ these definitions to account for the non-colluding servers model employed in several works\;\cite{abspoel2020secure,fang2009new,fang2010preserving,li2017outsourcing,li2017privacy,liu2020towards,ma2019privacy}. A \emph{collective} is a group of parties interested in training a tree-based machine-learning model on their joint \emph{global} dataset.

\section{Scope and Method}\label{sec:method}
We systematize the research efforts on privacy-preserving collaborative learning of decision-tree models in a thorough and structured manner. Overall, our focus is on the perspective of privacy-enhancing technologies (PETs). Hence, our goal is to understand their use for tree-based model induction algorithms, their application to the distributed setting, their trust assumptions, the challenges that they are confronted with, and their limitations and bottlenecks. Thus, we survey the current literature and analyze it from various viewpoints. In this section, we describe the methodology that we employed when searching the literature and the systematization approach that we devised to classify the relevant works.

\descr{Search Methodology.} We used Google Scholar\;\cite{scholar}, Microsoft Academic\;\cite{microsoft}, and DBLP\;\cite{dblp}, to discover works related to privacy-preserving collaborative tree-based model learning: Our search results comprised 73 papers from a wide range of research communities (see Appendix~\ref{app:analysis}). 
We cross-referenced each paper to discover additional relevant works. Overall, our search resulted in 103 papers about privacy-preserving collaborative learning of decision trees that we analyzed.

\descr{Systematization Methodology.} To classify and organize these works in a structured manner, we devise a systematization methodology that enables us to characterize their approaches on collaborative and privacy-preserving tree-based model induction. Our method takes into account four systematization axes that we briefly describe here:

\descrit{Learning Algorithm (see \S\ref{sec:algo}).} This refers to the techniques used for the tree-based model learning. These include the machine-learning task, the data type, the training algorithm, and the underlying quality metric. 

\descrit{Collaborative Model (see \S\ref{sec:collab}).} This axis is related to the entities involved in the training of the tree-based model, the computation, communication, and data model assumed: the actors involved, their roles, how they interact, and how the data is distributed among them.

\descrit{Protection Mechanism (see \S\ref{sec:ppm}).} The protection mechanism refers to the privacy-enhancing technologies employed to protect the different components that interact during the tree-based learning, \ie the training data, the intermediate values, and the final model weights. 

\descrit{Threat Model (see \S\ref{sec:tm}).} This systematization criterion is related to the definition of the capabilities and objectives of the adversaries that each work considers. It refers to the assumptions imposed to reach a target protection level for each component of the model learning.

\section{Decision-Tree Algorithms}\label{sec:dt}
We present various tree-based learning algorithms. 
Hereinafter, we adopt the following notation. We denote by $\mathcal{D}$ the global dataset of $n$ training samples $(\bm{x}, y)$, where $\bm{x}$ consists of $d$ features and $y{\in}C$ denotes its associated label, with $C$ the set of possible classes. For simplicity, we describe the notation and algorithms for classification tasks. Let $\mathcal{A}$ be the attribute set defined as $\mathcal{A}{=}\{A_1,{\dots}, A_d\}$. Each component $A_k {\in}\mathcal{A}$ is a set indicating the possible values obtained by the $k$-th feature of a sample. With a slight notation abuse, $A_k{=}v$ indicates that the $k$-th feature takes the value $v{\in} A_k$. For any set $S$, $|S|$ denotes its cardinality.

\subsection{Background on Decision Trees}\label{sec:dt:tree}
Decision-tree algorithms are tree-based supervised learning techniques used for classification or regression tasks. The structure of a tree can be seen as a directed acyclic graph that consists of nodes and edges. 
The \emph{root} node has no parent edges, and subsequent nodes are called \emph{internal}. If an internal node has no children, it is called a \emph{leaf}. During learning, let $D_i$ be the part of the dataset that reaches a node $i$ and that is split among its children. This decision is made depending on $D_i$'s value for a specific feature called a \emph{split point}. 
The learning (or induction) process of a tree determines the splitting feature and split point for every tree node and results in what is known as the tree's \emph{structure}.

\subsection{Greedy Algorithms}\label{sec:dt:greedy}
Classic decision-tree induction techniques rely on greedy algorithms\;\cite{quinlan1986induction,breiman1984classification}. These algorithms follow a top-down approach, \ie\,they build the tree from the root to the leaves. At each node, they attempt to find the optimal ``short-term'' decision. Thus, for each node, the learning problem is reduced to maximizing an objective function that depends on the algorithm.

\descr{Quality Metrics.} Among the most popular metrics for the objective function are the entropy-based information gain, the gain ratio, and the Gini index. 

\descrit{Entropy-Based Information Gain.} The entropy of the $i$-th node over the class labels is defined as
\begin{equation}
\text{H}_{C}(D_i){=} -\sum_{c\in C} \frac{|D_i^{c}|}{|D_i|}\cdot\log_2\left(\frac{|D_i^{c}|}{|D_i|}\right),
\end{equation}
with $|D_i^c|$ the number of samples in $D_i$ with class label $c \in C$.
The best split is defined as the partitioning of the dataset $D_i$ along the $k$-th feature that maximizes the information gain $\text{Gain}(k,D_i){=} \text{H}_C(D_i) {-} \text{H}_C(D_i|k)$, where $\text{H}_C(D_i|k)$ is the conditional information given by the $k$-th feature.

\descrit{Information-Gain Ratio.} For the $k$-th feature, this metric is the ratio between the gain $\text{Gain}(k,D_i)$ and the information value of the split defined as $\text{H}_{A_k}(D_i)$. This metric can be extended to a continuous feature $a$ by splitting its space in two parts (\ie $a{\leq} v$ and $a {>} v$)\;\cite{xiao2006privacy}.

\descrit{Gini Index.} Informally, it corresponds to the probability of incorrectly classifying a sample when randomly picking the label following the class value distribution for a specific feature. The Gini impurity metric is defined by replacing the entropy $\text{H}_C(\cdot)$ by the Gini function defined as $\text{Gini}(D_i){=}  1{-}\sum_{c\in C} \left(|D_i^c|/|D_i|\right)^2$.
This metric is more arithmetically-friendly than the previous two, as it can be rewritten with only additions and multiplications.
Similarly to the information gain, the best split of $D_i$ is along the feature that maximizes the Gain value.

\descrit{Other Splitting Techniques.} Some works explore alternatives to these classic splitting techniques. The median-based splitting criterion favours a balanced distribution of the data across leaves\;\cite{consul2020differentially}. The max operator corresponds to the mis-classification rate if the majority class is returned for a specific feature value\;\cite{zhu2013effective,patil2014differential,liu2018differentially,friedman2010data}.

\descr{Induction Algorithms.} There exist several algorithms that use the above metrics to build trees.
The ID3 algorithm\;\cite{quinlan1986induction} builds classification trees over categorical data. Following a top-down approach, and recursively at each node, the best split is computed on the dataset that reaches the node and the available features by using the information gain metric. All possible values of the selected feature are added as children, and the dataset is split accordingly. The process is repeated for every child until all dataset samples belong to the same class or until the feature set is empty (the majority class is returned). The CART algorithm\;\cite{breiman1984classification} accounts for both regression and classification tasks by using the Gini impurity metric. After generating the tree, it prunes it into a smaller one. The C4.5 algorithm\;\cite{quinlan1993c4} improves ID3 by using the gain ratio to handle both numerical and categorical data.

\subsection{Random Algorithms}\label{sec:dt:rand}
Contrary to the greedy algorithms above, random approaches generate a tree at random instead of using heuristics: For each node, a random feature is picked from the set of available ones, the tree is split based on all its possible values, and the selected feature is removed from the set. 
The structure of the tree is determined randomly beforehand, and the training data is used to prune the tree and label the leaves. This method is known as completely random trees\;(CRTs)\;\cite{fan2003random}. Geurts\;\etal\,\cite{geurts2006extremely} proposed extremely random trees (ExRTs) similar to CRTs but each node evaluates $K$ random splits and selects the one returning the best information gain.

\subsection{Random Forests}\label{sec:forest}
Breiman proposed the combination of multiple decision trees to form a random forest (RF)\;\cite{breiman2001random}. A forest is an ensemble of classification (or regression) trees trained independently. Each tree is trained with one of the aforementioned induction algorithms, using a set of bootstrap training samples and a random subset of features. Forest predictions are made by aggregating all the trees evaluations. We denote by CRF a RF made of CRTs.

\subsection{Boosting}\label{sec:dt:boost}
\descr{Adaptive Boosting\,(AdaBoost)\,\cite{freund1995desicion}.} 
AdaBoost works sequentially by adding simple models (classifiers or regressors) to the ensemble model with a certain weight: The final model is a weighted aggregate of these weak learners whose weight is decreasing according to their error. At each iteration, a model is trained focusing on mis-classified samples from the previous iteration. 

\descr{Gradient-Boosting Decision Tree\,(GBDT)\;\cite{friedman2001greedy}.} GBDT also follows a sequential approach: At each iteration, the updated model learns from the previous one. This is achieved by training a simple tree model on the residuals, \ie the difference between the observation and the output, of the previous tree. In GBDT, residuals for a datapoint $(\bm{x},y)$ are approximated at each iteration as the negative gradient of an objective function $L(y,f(\cdot))$ evaluated on $\bm{x}$ with $f(\cdot)$ the sum of classifiers from previous iterations. 
We refer the reader to Friedman's work\;\cite{friedman2001greedy} for further details.

\descr{XGBoost\,\cite{chen2016xgboost}.} This is an improvement of GBDT and currently the state-of-the-art tree-based algorithm\;\cite{sandulescu2016predicting}. Its main characteristic is the use of the second-order derivatives and a better regularization. XGBoost creates an ensemble of $K$ CARTs and its objective function is $L {=} \sum_{i=1}^n l(\hat{y}_i,y_i) {+} \sum_m\Omega(f_m)$, 
where $l(\cdot,\cdot)$ is a convex loss-function measuring the difference between the target label $y_i$ and the prediction $\hat{y}_i$, $f_m$ the $m$-th decision tree with $m \in \{1, \dots, K\}$, and $\Omega(f)$ a regularisation term. 
At each iteration $t$, the simplified objective is $
\tilde{L}^{(t)}{=}\sum_{i=1}^{n}[g_if_t(\bm{x}_i)+\frac{1}{2}f_t^2(\bm{x}_i)] + \Omega(f_t)$ where $f_t$ is the $t$-th iteration CART tree, $g_i$ and $h_i$ are the first- and second-order derivative of $l(y_i,\cdot)$, respectively. Then, representing a tree $f_k$ as a fixed structure with $T$ leaves of weights $w$, $\lambda$ and $\gamma$ parameters, and for a partition $I$ of data reaching leaf $j$, the optimal weight of $j$ is $w_j^* {=} - (\sum_{i \in I_j} g_i)/(\sum_{i \in I_j} g_i + \lambda)$.
The optimal tree structure is obtained with a greedy algorithm that finds the split maximizing the objective $L_{\text{split}}$. Denoting by $I_L$ and $I_R$ the instances of the left and right nodes, respectively, (\ie $I{=}I_L \cup I_R $), the split objective is
\begin{equation}
L_{\text{split}}(I){=} \frac{1}{2}\left(\frac{(\sum_{i\in I_L} g_i)^2}{\sum_{i\in I_L} h_i{+}\lambda}  {+} \frac{(\sum_{i\in I_R} g_i)^2}{\sum_{i\in I_R} h_i{+}\lambda} {-} \frac{(\sum_{i\in I} g_i)^2}{\sum_{i\in I} h_i{+}\lambda} \right) {-}\gamma.
\end{equation}

\section{Learning Algorithm}\label{sec:algo}
As presented in \S\ref{sec:dt}, there exist numerous algorithms that can be used to learn tree-based models. Among these, ID3 is predominantly considered for the case of privacy-preserving collaborative decision trees, with \emph{classification} being the most popular task (see Table\;\ref{tab:algo}). Hence, most of the surveyed works only consider \emph{categorical} data, whereas some of them propose the discretization of continuous data to account for \emph{numerical} attributes\;\cite{fletcher2015a}. We classify as ``ensemble'' any work that creates a forest rather than single trees. For optimization reasons, several works modify the quality metric used by the learning algorithm\;\cite{abspoel2020secure,behera2011privacy,consul2020differentially,Hoogh2014practical,fang2009new,fang2010preserving,liu2009privacy,liu2018differentially,ma2008secure,samet2008privacy,wu2016privately,xin2019differentially} (see \S\ref{sec:ppm}). Overall in the literature, we observe a wide range of combinations on tree-types, tasks, data, and algorithms, as illustrated on Table\;\ref{tab:algo}. We remark that the learning algorithm imposes constraints on designing a privacy-preserving and collaborative solution as we will detail this issue in subsequent sections.

\begin{table}[t!]
\caption{\small{Learning algorithms in the surveyed literature.
\scriptsize{(Ens.: Ensemble, Clas.:Classification, Reg.:Regression, Num.:Numerical, Cat.:Categorical)}.}}\vspace{-0.2cm}
\label{tab:algo}
\fontsize{7}{7.2}\selectfont
\centering
\begin{tabularx}{0.49\textwidth}{m{3.9cm} >{\centering\arraybackslash\columncolor[gray]{0.9}}p{0.2cm} p{0.2cm} >{\columncolor[gray]{0.9}}p{0.2cm} p{0.2cm} >{\columncolor[gray]{0.9}}p{0.2cm} l}
\toprule
& \multicolumn{1}{c}{}& \multicolumn{2}{c}{Task}& \multicolumn{2}{c}{Data}& \\
\cmidrule(lr){3-4} \cmidrule(lr){5-6}
\multicolumn{1}{c}{Reference} & \rot{\small{Ens.}}&\rot{Clas.} &\rot{Reg.} &\rot{Num.} &\rot{Cat.} &\multicolumn{1}{c}{Algorithm}\\
\midrule
\cite{wu20privacy,xiang2018collaborative} & \pie{360} & \pie{360} & \pie{360} & \pie{360} & \pie{360} & CART \\ \hdashline
\cite{dansana2013novel} && \pie{360} & \pie{360} & \pie{360} & \pie{360} & CART \\ \hdashline
\cite{consul2020differentially} && \pie{360} & \pie{360} & \pie{360} & \pie{360} & CART-like\\ \hdashline
\cite{abspoel2020secure} && \pie{360} & \pie{360} & \pie{360} & \pie{360} & C4.5-like\\ \hdashline
\cite{vaidya2013random} && \pie{360} & \pie{360} & \pie{360} & \pie{360} & CRT\\ \hdashline
\cite{liu2020federatedExtraTree} && \pie{360} & \pie{360} & \pie{360} & \pie{360} & ExRT\\ \hdashline
\cite{liu2019revocable} && \pie{360} & \pie{360} & \pie{360} && ExRT\\ \hdashline
\cite{fletcher2015a} && \pie{360} & \pie{360} && \pie{360} & CART \\ \hdashline
\cite{baghel2013privacy,behera2011privacy,gangrade2009building,kadampur2010noise,kikuchi2013privacy,li2019outsourced,sharma2016secure,shen2009privacy,Tandel2016PrivacyPD,xiao2006privacy} && \pie{360} && \pie{360} & \pie{360} & C4.5\\ \hdashline
\cite{du2002building} && \pie{360} && \pie{360} & \pie{360} & CART \\ \hdashline
\cite{wang2006classification,zhu2013effective} && \pie{360} && \pie{360} & \pie{360} & CART-like\\ \hdashline
\cite{sumalatha2016fuzzy} && \pie{360} && \pie{360} & \pie{360} & CRT\\ \hdashline
\cite{friedman2010data,wang2020scalable} && \pie{360} && \pie{360} & \pie{360} & ID3/C4.5 \\ \hdashline
\cite{liu2009privacy} && \pie{360} && \pie{360} && C45\\ \hdashline
\cite{agrawal2000privacy,akavia2019privacy,bu2007preservation} && \pie{360} && \pie{360} && CART \\ \hdashline
\cite{alabdulkarim2019ppsdt} && \pie{360} &&& \pie{360} & C4.5\\ \hdashline
\cite{bai2017embedding,brickell2009privacy,guan2020differentially,xin2019differentially} && \pie{360} &&& \pie{360} & CART \\ \hdashline
\cite{khodaparast2018privacy,li2017outsourcedSensor,wu2020differentially} && \pie{360} &&& \pie{360} & CRT\\ \hdashline
\cite{jagannathan2009practical} && \pie{360} &&& \pie{360} & ID3/CRT\\ \hdashline
\cite{bojarski2014differentially} & \pie{360} & \pie{360} & \pie{360} & \pie{360} & \pie{360} & CRF\\ \hdashline
\cite{zhao2018inprivate} & \pie{360} & \pie{360} & \pie{360} & \pie{360} & \pie{360} & GBDT \\ \hdashline
\cite{giacomelli2019privacy} & \pie{360} & \pie{360} & \pie{360} & \pie{360} & \pie{360} & RF \\ \hdashline
\cite{liu2020federated} & \pie{360} & \pie{360} & \pie{360} & \pie{360} & \pie{360} & RF/CART\\ \hdashline
\cite{li2019practical,li2020privacy} & \pie{360} & \pie{360} & \pie{360} & \pie{360} && GBDT \\ \hdashline
\cite{cheng2019secureboost,law2020secure,leung2019towards,liu2019boosting,tian2020federboost,wang2020cloud,yang2019tradeoff} & \pie{360} & \pie{360} & \pie{360} & \pie{360} && XGBoost\\ \hdashline
\cite{aslett2015encrypted,fletcher2017differentially} & \pie{360} & \pie{360} && \pie{360} & \pie{360} & CRF\\ \hdashline
\cite{hou2019dprf,li2020multicenter,patil2014differential,rana2015differentially} & \pie{360} & \pie{360} && \pie{360} & \pie{360} & RF \\ \hdashline
\cite{gambs2007privacy,wang2020privacypreserving} & \pie{360} & \pie{360} && \pie{360} && AdaBoost \\ \hdashline
\cite{ma2019privacy} & \pie{360} & \pie{360} && \pie{360} && RF \\ \hdashline
\cite{fang2020hybrid,feng2019securegbm} & \pie{360} & \pie{360} && \pie{360} && XGBoost\\ \hdashline
\cite{nock2020boosted} & \pie{360} & \pie{360} &&& \pie{360} & Boosted\\ \hdashline
\cite{fletcher2015differentially} & \pie{360} & \pie{360} &&& \pie{360} & CRF\\ \hdashline
\cite{zhang2017dpets} & \pie{360} & \pie{360} &&& \pie{360} & ExRT\\ \hdashline
\cite{alabdulkarim2019privacy} & \pie{360} & \pie{360} &&& \pie{360} & RF \\ \hdashline
\cite{blum2005practical,du2003using,emekci2007privacy,fang2008privacy,fang2009new,fang2010preserving,fong2010privacy,gangrade2011novel,gangrade2012privacy,giannella2004communication,han2007multi,Hoogh2014practical,kalyani2017decision,kuijpers2008privacy,li2017outsourcing,li2017privacy,lindell2000privacy,liu2018differentially,liu2020towards,lory2012enhancing,ma2008secure,rao2014cryptographic,samet2008privacy,sheela2013novel,suthampan2005privacy,teng2007hybrid,truex2019hybrid,vaidya2005privacy,vaidya2008privacy,wu2020differentially,xiao2005privacy,zhan2005privacy,zhan2007using} && \pie{360} &&& \pie{360} & ID3\\ 
\bottomrule
\end{tabularx}
\vspace{-0.3cm}
\end{table}

\section{Collaborative Model}\label{sec:collab}

We propose a systematization of the literature on privacy-preserving collaborative tree-based model learning based on their collaborative model. In distributed settings, it is crucial to understand which entities are involved, their role, and how they interact. We distinguish two major aspects of the collaborative model: (i) the computation and communication model and (ii) the data distribution model. The former considers the entities and their interactions, whereas the latter describes how the global dataset is partitioned.

\subsection{Computation and Communication Model}\label{sec:collab:fl}

We consider three types of entities that participate in the distributed learning: parties, miner(s), and aggregator (see \S\ref{sec:intro:not}). Although some works rely on additional (external) entities (\eg a trusted third-party, a public-key infrastructure), here we omit them as they do not directly participate in the learning process. 
We present the different collaborative models considered in the literature on distributed privacy-preserving tree-based model induction. The different categorization corresponds to where the training data is available and where the bulk of the computation is executed.

\descr{Central and Offloading Model.} In this case, a miner has access to all the parties' data and performs the training. This model covers works that are not initially envisioned for the distributed setting, \eg\cite{blum2005practical,brickell2009privacy,consul2020differentially,fletcher2015a,friedman2010data,jagannathan2009practical,li2020privacy,liu2018differentially,patil2014differential,xin2019differentially,zhu2013effective}, as well as works under the \emph{offloading} category where the parties explicitly communicate their dataset to the miner, \eg\cite{agrawal2000privacy,akavia2019privacy,aslett2015encrypted,brickell2009privacy,estivill1999data,fong2010privacy,ma2019privacy}. 
Brickell and Shmatikov propose a model where a miner creates a private model on a single remote secret database\;\cite{brickell2009privacy}.
For privacy, Abspoel\;\etal assume parties offload their data by sharing it across three non-colluding servers\;\cite{abspoel2020secure}. The offloading to only two servers is also possible\;\cite{ma2019privacy}.

\descr{Aggregator Model.} Contrary to the previous model, in this one, any party that takes part in the learning process communicates with an aggregator. The latter coordinates the training by obtaining intermediate values computed locally by the parties on their data. The aggregator combines these values and publishes the result such that the learning process can continue. The aggregator operates differently, depending on the learning task. For example, it issues count queries to the parties to compute the information gain for the ID3 algorithm (\eg\cite{truex2019hybrid}) or selects a subset of features to be considered by parties (\eg\cite{liu2020federatedExtraTree}). 
In another approach, trees are learned locally and sent encrypted to the aggregator (\eg\cite{giacomelli2019privacy}). The aggregator uses all the locally trained encrypted trees to compute the encrypted prediction. But, experiments show a performance drop in the model accuracy\;\cite{giacomelli2019privacy}. A similar approach is also used for distributed  AdaBoost\;\cite{li2016distributed,xiang2018collaborative}.

\descr{Fully Distributed (FD) Model.} This scenario involves only members of the collective that hold local data. 
In this case, each party communicates with all the others. Although initially considered for the setting of two parties by Lindell and Pinkas\;\cite{lindell2000privacy} and subsequent works \cite{du2002building,fang2020hybrid,feng2019securegbm,giannella2004communication,teng2007hybrid,xiao2006privacy}, it was later extended to arbitrary number of parties \cite{emekci2007privacy,gambs2007privacy,Hoogh2014practical,leung2019towards,li2016distributed,li2019practical,samet2008privacy,xiao2005privacy,kuijpers2008privacy,vaidya2013random,wu20privacy}. In this setting, some parties might be assigned specific tasks. We refer to systems with consistent parties' roles during learning as \emph{leader}-based \cite{cheng2019secureboost,han2007multi,sheela2013novel,sumalatha2016fuzzy,zhan2005privacy,zhan2007using}, whereas those with temporary ones as \emph{sequential}\;\cite{zhao2018inprivate}: \eg the local learning of a model before communicating it to the next member of the collective.

\subsection{Data-Distribution Model}\label{sec:collab:data}
The global dataset can be partitioned in different ways among the entities that participate in the tree-based model-learning process:

\descr{No Partitioning.} The whole data is centralised in one unique site and the learning is executed on it. It is directly related to the central and offloading models presented in \S\ref{sec:collab:fl} and often a comparison baseline. Although it requires parties to communicate their data to a remote server, no subsequent rounds of communication are needed as the learning operates as if the data was never distributed.

\descr{Horizontal Partitioning.} All the parties possess their own data samples that share the same feature space. The collaborative learning incurs communication overhead as the parties need to exchange intermediate results computed on their local data. For instance, for the ID3 algorithm, the collective needs to compute the information gain for every possible split: This requires sharing the counts of the data points at each split. This becomes communication heavy, with an increasing number of features and classes, hence multiple works modify the learning algorithm to a randomized one that is more collaboration-friendly; indeed, only the final leaf counts need to be computed on the distributed data.

\descr{Vertical Partitioning.} Conversely to the previous case, with vertical partitioning, the parties share a different feature space for the same individual samples of the global dataset. This can, for instance, represent a distributed database containing the data of a common set of customers across multiple financial institutions. The class label is known by all the parties or by only one. The challenge in this setting is to find the feature on which to split the tree. With greedy algorithms, to find the feature with the best gain, each party can locally compute the gain, and then communicate with others.

\descr{Arbitrary Partitioning.} In this case, both horizontal and vertical partitioning are present. It is often referred to as an arbitrary (or mixed) partitioning of the data. Only a few works present a distributed solution for this case by using special data representation\;\cite{han2007multi} or by adding extra rounds of communication\;\cite{Hoogh2014practical,kuijpers2008privacy,ma2008secure,vaidya2013random}.

\subsection{Summary}
Table\;\ref{tab:model} presents the collaborative models of the surveyed works. The vast majority of works consider the aggregator or offloading model. Works in the central model can trivially handle both data partitioning tasks, as they eventually gather the data in one place. The fully distributed case is by far the most challenging one. However, most of those solutions focus on the two-party case. Vertical and horizontal partitioning have, respectively, their own limitations: The former relies on local best attributes but often needs public class labels, and the latter requires distributed computations, introducing computation and communication overhead. 
Few works acknowledge the challenges introduced by the local data following different statistical distributions\;\cite{fletcher2019decision,li2020multicenter,li2019practical,li2016distributed,zhao2018inprivate}. Although this is a non-issue for greedy algorithms in which the distributed algorithm works as if the data was centralised, it can become cumbersome for randomized and boosted algorithms\;\cite{li2020multicenter}.
Moreover, the notion of availability or dropout of the parties is often overlooked with only one work that provides a solution for parties to leave the collective and that removes their impact on the training\;\cite{liu2019revocable}.

\begin{table}
\caption{\small{Different collaborative models in the literature.}}
\label{tab:model}
\fontsize{7}{7.2}\selectfont
\centering
\begin{tabularx}{0.49\textwidth}{m{3.3cm}l>{\centering\arraybackslash\columncolor[gray]{0.9}}cc>{\columncolor[gray]{0.9}}cc>{\columncolor[gray]{0.9}}cc}
\toprule
& & \multicolumn{3}{c}{Members}& \multicolumn{3}{c}{Data Model} \\
\cmidrule(lr){3-5} \cmidrule(lr){6-8}
\multicolumn{1}{c}{Reference} & \multicolumn{1}{l}{Coll.
Model} & \rot{\scriptsize{Parties}} & \rot{\scriptsize{Aggregator}} & \rot{\scriptsize{\;Miner}} & \rot{\scriptsize{Vertical}} & \rot{\scriptsize{Horizontal}} & \rot{\scriptsize{Both}} \\[-0.3em]
\midrule
\cite{Hoogh2014practical,kuijpers2008privacy,ma2008secure,vaidya2013random}& FD& \pie{360} &&&&& \pie{360}\\ \hdashline
\cite{gangrade2009building,suthampan2005privacy,vaidya2008privacy,wu20privacy} & FD& \pie{360} &&& \pie{360} && \\ \hdashline
\cite{behera2011privacy,emekci2007privacy,gambs2007privacy,leung2019towards,samet2008privacy,xiao2005privacy}& FD& \pie{360} &&&& \pie{360} & \\ \hdashline 
\cite{khodaparast2018privacy}& FD \tiny{(2 parties)}& \pie{360} &&&&& \pie{360}\\ \hdashline
\cite{du2002building,fang2020hybrid,feng2019securegbm,giannella2004communication,kikuchi2013privacy,teng2007hybrid}& FD \tiny{(2 parties)}& \pie{360} &&& \pie{360} && \\ \hdashline
\cite{lindell2000privacy,lory2012enhancing,xiao2006privacy}& FD \tiny{(2 parties)}& \pie{360} &&&& \pie{360} & \\ \hdashline
\cite{han2007multi}& Leader& \pie{360} &&&&& \pie{360}\\ \hdashline
\cite{cheng2019secureboost,dansana2013novel,sheela2013novel,vaidya2005privacy,zhan2007using} & Leader& \pie{360} &&& \pie{360} && \\ \hdashline
\cite{gangrade2011novel,li2019practical,sumalatha2016fuzzy,zhan2005privacy}& Leader& \pie{360} &&&& \pie{360} & \\ \hdashline
\cite{zhao2018inprivate} & Sequential& \pie{360} &&&& \pie{360} & \\ \midrule
\cite{li2019outsourced,tian2020federboost}& Aggregator& \pie{360} & \pie{360} &&&& \pie{360}\\ \hdashline
\cite{fang2008privacy,liu2019revocable,liu2020federated,sharma2016secure,shen2009privacy,wang2006classification} & Aggregator& \pie{360} & \pie{360} && \pie{360} && \\ \hdashline 
\cite{alabdulkarim2019ppsdt,alabdulkarim2019privacy,gangrade2012privacy,giacomelli2019privacy,law2020secure,li2017outsourcedSensor,li2020multicenter,liu2019boosting,liu2020federatedExtraTree,truex2019hybrid,wang2020cloud,xiang2018collaborative,yang2019tradeoff}& Aggregator& \pie{360} &&&& \pie{360} & \\ \midrule
\cite{abspoel2020secure,fang2010preserving,liu2020towards,ma2019privacy} & Offloading& \pie{360} & \pie{360} & \pie{360} &&& \pie{360}\\ \hdashline
\cite{fang2009new,li2017outsourcing,li2017privacy} & Offloading& \pie{360} & \pie{360} & \pie{360} && \pie{360} & \\ \hdashline
\cite{agrawal2000privacy,akavia2019privacy,aslett2015encrypted,baghel2013privacy,bu2007preservation,du2003using,fong2010privacy,kadampur2010noise,kalyani2017decision,liu2009privacy,mohammed2011differentially,abspoel2020secure,rao2014cryptographic,Tandel2016PrivacyPD,wang2020privacypreserving}& Offloading& \pie{360} && \pie{360} &&& \pie{360}\\ \hdashline 
\cite{bai2017embedding,blum2005practical,bojarski2014differentially,brickell2009privacy,consul2020differentially,fletcher2015a,fletcher2015differentially,fletcher2017differentially,friedman2010data,guan2020differentially,hou2019dprf,jagannathan2009practical,li2020privacy,liu2018differentially,nock2020boosted,patil2014differential,rana2015differentially,xin2019differentially,zhang2017dpets,zhu2013effective,wang2020scalable} & Central & \pie{360} && \pie{360} &&& \pie{360} \\
\bottomrule
\end{tabularx}
\vspace{-0.3cm}
\end{table}

\section{Protection Mechanism}\label{sec:ppm}
We review the types of privacy-enhancing technologies (PETs) employed to ensure confidentiality of data during the collaborative tree induction process. We identify five categories of PETs: (a)\,input randomization\,(\S\ref{sec:ppm:perturb}), (b)\,differential privacy-based solutions\,(\S\ref{sec:ppm:DP}), (c)\,cryptographic approaches (\S\ref{sec:ppm:smc}), (d)\,hardware-based solutions\,(\S\ref{sec:ppm:hw}), and (e)\,hybrid solutions that combine the above\,(\S\ref{sec:ppm:hybrid}).

\subsection{Input Perturbation and Randomization}\label{sec:ppm:perturb}
These techniques generate a surrogate dataset by perturbing the original one to protect its confidentiality; this dataset is used by the miner to train the tree-based model. Thus, it is predominantly employed in the offloading collaborative model, \eg\cite{agrawal2000privacy,dowd2006privacy,bu2007preservation,estivill1999data,fong2010privacy,kadampur2010noise,liu2009privacy}.
Typically, perturbation techniques discretize or add noise to each sensitive attribute of the dataset\;\cite{agrawal2000privacy}, or swap the dataset with a surrogate one that has the same probability distribution\;\cite{estivill1999data}. However, Kargupta\;\etal\,\cite{kargupta2003privacy} showed that noise addition does not prevent the reconstruction of patterns in the data: They propose a filtering technique that reconstructs the original dataset. Whereas other random substitution techniques are immune to such attacks\;\cite{dowd2006privacy}, they hamper utility as the mining is performed on an approximation of the original data\;\cite{fong2010privacy}. Some works reconstruct a surrogate dataset from the sanitized original database\;\cite{kalyani2017decision}, whereas others assume that the miner obtains the tree computed on the original data before adding noise\;\cite{kadampur2010noise}. Finally, some studies modify the learning algorithm to handle the perturbed datasets\;\cite{liu2009privacy}.

Although this line of work does not necessarily target the distributed scenario, it can be trivially extended to it. Overall, input perturbation and randomization techniques are ad-hoc methods that obfuscate the training dataset with limited privacy guarantees. So, more recent works rely on differential privacy to address this lack of formalization.

\subsection{Differential Privacy Based Solutions}\label{sec:ppm:DP}
Introduced by Cynthia Dwork\;\cite{dwork}, differential privacy (DP) is a privacy definition for sharing information with respect to queries on a database. Informally, it guarantees that the change of a single data record does not modify the query's output significantly. For a query $f$, this is achieved by adding noise to its answer; the noise amount is determined by the query's sensitivity $\Delta(f)$ and the privacy budget $\varepsilon$.
When designing a DP-based solution for collaborative decision-tree learning, four important aspects should be considered: which entity performs the noise addition, at which training stage, the magnitude of noise required, and the total privacy budget spent for training. Tackling these enables to train and publish trees with DP guarantees.

\descr{Central Model.}
Recall that, in this scenario, a single entity has access to the entire training dataset. 
The main idea is to inject noise during key learning parts, \eg for selecting the best feature\;\cite{friedman2010data}, counting class counts at the leaves\;\cite{jagannathan2009practical,consul2020differentially}, or computing gain queries for each feature\;\cite{fletcher2015a}. More recent works aim to find tighter sensitivity bounds for the training queries or new ways to embed DP\;\cite{bai2017embedding}.
Other approaches \textit{relax} the learning algorithm by replacing information gain with more DP-friendly metrics, \eg Gini\;\cite{fletcher2015a} or Max\;\cite{liu2018differentially}. For instance, the max operator has lower sensitivity than the Gini or information gains thus leading up to higher accuracy on similar datasets and privacy levels\;\cite{friedman2010data,fletcher2019decision}.
In other works, the learning is adapted using RFs\;\cite{hou2019dprf,patil2014differential,rana2015differentially} or CRTs\;\cite{bojarski2014differentially,fletcher2015differentially,fletcher2017differentially,zhang2017dpets}.
Some works abusively consider that each tree in the forest is independent, to reduce the privacy budget consumption\;\cite{rana2015differentially}. However, this is circumvented by training each tree on an independent subset of the training data and applying the parallel composition theorem\;\cite{fletcher2015differentially,xin2019differentially}.
Overall, works based on the central model consider reasonable privacy budgets (\ie $\varepsilon{\in}[0.1;1.0]$) and some even experiment with very low budgets (\eg $\varepsilon{=}0.01$ for\;\cite{bai2017embedding,bojarski2014differentially,fletcher2015differentially,liu2018differentially,nock2020boosted}). We note that the privacy budget configuration directly affects the model's accuracy; Fletcher\;\etal report accuracy drops of more than $20$\% when the budget is reduced from $2.0$ to $0.2$ to obtain stronger privacy guarantees\;\cite{fletcher2017differentially}. We refer the interested reader to the survey by Fletcher and Islam on differentially private trees\;\cite{fletcher2019decision}, as it exclusively focuses on the central model where the miner has access to the dataset. 

\descr{Aggregator Model.}
In this setting, parties apply perturbations to their local intermediate results before sharing them with the potentially untrusted aggregator. Xiang\;\etal\,\cite{xiang2018collaborative} present two collaborative algorithms: CART-based random forest and AdaBoost. 
In their solution, each party builds a local ensemble model by injecting noise into each tree, and the aggregator merges the perturbed models with a weighted sum that depends on each party's data samples and reported accuracy. Similarly, Liu\;\etal build a differentially private ExRT where the split is randomly selected to reduce the privacy budget consumption\;\cite{liu2020federatedExtraTree}. 
In the work by Li\;\etal\;\cite{li2020multicenter}, each party uses a differentially-private Generative Adversarial Network\;(GAN) to generate a surrogate ``synthetic'' dataset. This can be shared with the aggregator; it merges the received datasets and re-distributes them to the parties that use them as validation data to select the best local model. Finally, the best local models are shared with the aggregator that combines them into the final global model.

\descr{Fully Distributed Model.} Zhao\;\etal\,\cite{zhao2018inprivate} propose a collaborative system for GBDT. They employ an iterative sequential method where each party locally trains on its data and transfers the resulting tree to the next party. To ensure privacy of the shared model, each party generates DP trees: The split values are sampled using the exponential mechanism~\cite{mcsherry2007mechanism}.

\subsection{Cryptographic Solutions}\label{sec:ppm:smc}
Numerous works employ cryptographic techniques to protect the confidentiality of the tree-based model induction process. The most common cryptographic tools are Secure Multiparty Computation\;(SMC) (\ie secret sharing, garbled circuits, and oblivious transfers), homomorphic encryption (HE), encryption as obfuscation, and locality-sensitive hashing. We provide a brief background on these before analyzing their use by the relevant works. Table\;\ref{tab:leak} displays the different cryptographic solutions employed in the literature (see PET column).

\descr{Background.} Secret sharing\;(SS) methods distribute a secret into \emph{shares} such that the secret can only be reconstructed by re-combining all the shares (or a subset of them piloted by a threshold). Oblivious transfer\;(OT)\;\cite{naor1999oblivious}, oblivious polynomial evaluation\;\cite{naor2006oblivious}, and garbled circuits\;(GC)\;\cite{yao1986generate} are SMC building blocks that enable private secrets exchange, function computation and, overall, circuit evaluation on private inputs. Generic frameworks (\eg SPDZ\;\cite{damgaard2012multiparty,mp-spdz} or VIFF\;\cite{viff}) use these techniques and provide an abstraction for SMC supporting arithmetic operations, comparisons, and more. 
Other useful protocols include private set intersection\;(PSI)\;\cite{kissner2005privacy,pinkas2014faster,verykios2004state}, cardinality of intersections\;\cite{agrawal2003information,freedman2004efficient,vaidya2005privacy}, and secure scalar product\;(SSP)\;\cite{du2002building}; these protocols are at the core of some distributed tree induction algorithms: \ie counting the number of samples reaching a node using dot product between binary vectors\;\cite{du2002building} or creating consensus among two parties\;\cite{gambs2007privacy}. 
Homomorphic encryption\;(HE) enables computations on ciphertexts without requiring decryption. Depending on the scheme, operations can be linear\;(LHE)\;\cite{elgamal1985public,paillier1999public} or, with fully homomorphic\;(FHE) schemes, polynomial ones\;\cite{brakerski2014leveled,cheon2017homomorphic}.
HE can reduce the communication overhead of SMC: Instead of being shared, the secret is encrypted and computations are done directly on the ciphertexts. Also, as HE schemes support only limited operations, combining them with SMC in the offloading model enables new functionalities such as divisions or comparisons\;\cite{liu2016efficient}. Recent schemes merge directly HE and SMC for efficiency\;\cite{chen2019efficient,mouchet2020multiparty,kim2020how}. 
Additionally, HE and SS can be used to keep the final model secret and support oblivious predictions\;\cite{abspoel2020secure,aslett2015encrypted,giacomelli2019privacy,liu2019revocable,ma2019privacy,vaidya2013random}. 
We now present how SMC and HE techniques are used to protect distributed tree-based model induction by recognizing the constraints imposed by the learning and collaborative model chosen (\eg required information, communication topology, or data partitioning).

\descr{Secure Multiparty Computation.} Du and Zhan\,\cite{du2002building} pioneered the use of SMC for privacy-preserving ID3 training over a vertically distributed dataset among two parties. In their system, an external semi-honest entity generates blinding shares that are used, during the training process, for secure scalar-product operations: The intermediate counts are computed by dot products between binary vectors that represent the constraints needed to reach a specific node. Similarly, Lindell and Pinkas proposed the first algorithm for ID3 induction with horizontally distributed data across two parties\;\cite{lindell2000privacy}. Their algorithm uses GC techniques to obtain the attribute with the minimal conditional entropy and to compute an approximation of the $x\log x$ function required to calculate the information gain. However, it does not scale well with increasing number of parties. 
To this end, Emekçi\;\etal use Shamir's secret sharing\;\cite{shamir1979share} and propose a new secure-sum protocol to aggregate the counts required for the learning process\;\cite{emekci2007privacy}. Compared to their previous work\;\cite{emekci2006privacy}, they relax the need for a trusted aggregator and include a correctness check to thwart malicious parties that tamper with intermediate results during the computations. In particular, they increase the degree of the random polynomial used for the sharing, and they introduce redundancy in their system, which makes it computationally infeasible for a party to cheat. Nevertheless, their solution enables secure computation of only the gains; the best feature and the data split is performed in cleartext. 
Ma and Deng\,\cite{ma2008secure} reduce the communication and computation overhead of arbitrarily distributed ID3 by replacing the information gain with the Gini impurity metric. Privacy is ensured via secret sharing; multiplication, comparison, and addition operations are achieved with custom protocols executed among the parties. However, a trusted server is required to generate shares of secret blinding values. Gambs\;\etal\,\cite{gambs2007privacy} use secure-sum \,\cite{kantarcioglu2004privacy} and PSI\;\cite{kissner2005privacy} protocols to enable distributed AdaBoost. De Hoogh\;\etal\:\cite{Hoogh2014practical} opt for a generic framework\;\cite{viff} for SMC by using Shamir secret sharing: They train trees using the ID3 algorithm with the Gini index metric, but their solution supports only categorical data. This limitation was recently addressed by Abspoel\;\etal\,\cite{abspoel2020secure} by using SPDZ\;\cite{mp-spdz}.

Contrary to perturbation or DP-based approaches, SMC solutions enable almost \emph{exact} learning (to the approximation of non-polynomial operations, \eg ID3), hence they do not compromise the accuracy of the resulting model. Additionally, a handful of SMC solutions protect the resulting model by keeping it private (\ie secret-shared)\;\cite{abspoel2020secure,Hoogh2014practical}. SMC solutions also accommodate multiple parties. Their main drawback, however, is the introduced computation and communication overhead. Furthermore, to enable specific computations, \eg multiplications, there is sometimes the need for a trusted setup or a trusted server that generates intermediate values (\eg blinding shares or Beaver triples)\;\cite{du2002building,fang2020hybrid,liu2020towards,wu20privacy}.

\descr{Homomorphic Encryption.} Considered initially for offloading and outsourcing scenarios\;\cite{akavia2019privacy,aslett2015encrypted}, HE can also be used in collaborative settings by using the appropriate keying material. Most works that employ HE-based approaches use the additive scheme introduced by Paillier\;\cite{paillier1999public}, its threshold variant\;\cite{damgaard2001generalisation}, or similar multi-party schemes\;\cite{bresson2003simple}.
HE alleviates some limitations of pure SMC approaches, \eg\,the communication overhead and the need for a trusted setup. Indeed, with HE, secrets do not need to be shared among all parties, and computations can be executed by a single party without compromising privacy. On the negative side, HE schemes are limited by the operations allowed on ciphertexts; these schemes might not suffice to execute some tree-based learning algorithms.

Some works employ HE solely for the computation of the gain; the selection of the best attribute is computed in cleartext. Vaidya\;\etal\,\cite{vaidya2008privacy} introduce a secure dot product protocol using the Paillier cryptosystem to overcome the two-party limitation of prior work\;\cite{du2002building}. This protocol improves previous set-intersection algorithms\;\cite{agrawal2003information,freedman2004efficient,vaidya2005privacy}, by limiting leakage of unused information. Similar HE-based custom techniques for set-intersection cardinality or scalar product are also employed in the literature\;\cite{dansana2013novel,han2007multi,teng2007hybrid,liu2019revocable}. HE is also used for aggregating the local encrypted counts or statistics required to compute the gains\;\cite{alabdulkarim2019ppsdt,cheng2019secureboost,fang2009new,fang2010preserving}. Once aggregated, the result is decrypted and subsequent operations are conducted in cleartext. Several HE-based works opt for the Gini metric as it has an arithmetic representation simpler than the other metrics\;\cite{fang2009new,fang2010preserving,samet2008privacy}.

Other HE-based works modify the collaborative model. This enables the computation of functions not supported by LHE. In the leader collaborative model, Zhan\;\etal\,\cite{zhan2005privacy,zhan2007using}, propose solutions with three leading parties: One performs Paillier encryption/decryption and computations, a second generates randomness, and the third is in charge of blindings. This design enables the computation of non-linear functions and comparisons. However, the special-role parties need to be available and to follow the protocol. In the offloading scenario, some works introduce an additional entity\;\cite{li2017privacy,li2019outsourced,ma2019privacy}. This helps with the computation of non-linear functions and comparisons through multi-party HE protocols among two non-colluding parties\;\cite{bresson2003simple}. Sometimes parties directly assist with the computations\;\cite{akavia2019privacy}: The gain is computed on cleartext data, and the comparison is replaced by an approximation of the step function.

Finally, other HE-based works relax the learning algorithm and employ CRTs\;\cite{vaidya2013random,li2017outsourcedSensor,sumalatha2016fuzzy,khodaparast2018privacy}. The training data is used only for updating the leaves' statistics, thus HE can be used to gather these counts\;\cite{vaidya2013random}. Alternatively, each party creates a CRT or a local tree that is added in the global forest: HE-based consensus can be used to select which trees to retain\;\cite{li2017outsourcedSensor} or the trees are directly shared in encrypted form\;\cite{giacomelli2019privacy}.

Overall, HE is a powerful tool that can reduce the communication overhead of pure SMC solutions. Although it is affected by similar constraints to SMC (\eg requiring simple arithmetic circuit representations of the computations), HE solutions also introduce new challenges; the main one is the computation of the best information gain. 
Finally, we note that the adversarial model influences the choice of the HE scheme and its efficiency: \eg the widely used Paillier\;\cite{paillier1999public} or El-Gamal\;\cite{elgamal1985public} LHE schemes are not secure against quantum adversaries; recent lattice FHE schemes such as BGV\;\cite{brakerski2014leveled} or CKKS\;\cite{cheon2017homomorphic} alleviate this limitation at the cost of larger ciphertexts.
However, we found only a few works that use recent FHE schemes and focus on the offloading model\;\cite{akavia2019privacy,aslett2015encrypted}. They need a specific data representation of categorical data (one-hot-encoding or discretization) for equality tests and comparisons\;\cite{aslett2015encrypted}. 

\descr{Combining HE and SMC.} SMC and HE approaches are complementary. HE can reduce the communication overhead of SMC, whereas the latter supports arithmetic operations (\eg comparisons) that are inefficient with HE. Thus, many works on privacy-preserving collaborative decision-tree learning combine them to exploit the best of both worlds.
For greedy algorithms, HE can be used to compute aggregate intermediate values by combining encrypted local values of the $x\log x$ function, the Gini index, and the gain ratio. The subsequent comparisons to find the best attribute can be done privately using garbled circuits\;\cite{brickell2009privacy,xiao2006privacy}.  
Kikuchi\;\etal combine SMC\;\cite{du2002building} and HE\;\cite{vaidya2008privacy} approaches, and they design a secure scalar product protocol that incurs low communication costs and does not require a trusted setup\;\cite{kikuchi2013privacy}.
For XGBoost, several works use HE and secret sharing to protect the local intermediate residuals\;\cite{wang2020cloud, liu2019revocable,liu2019boosting}. Liu\;\etal\,\cite{liu2019boosting} propose an aggregation scheme that, with Shamir secret sharing and Paillier HE, ensures that the aggregator cannot access individual party updates. Each party locally computes gradients and the aggregator derives the score of each party's split to select in clear the best one.
Similarly, Fang\;\etal\,\cite{fang2020hybrid} propose a solution combining additive secret-sharing and HE. Contrary to similar works relying on pure HE\;\cite{cheng2019secureboost} or combined with hypervisor-enforced domain isolation and OT\;\cite{feng2019securegbm}, their solution maintains every value encrypted or secret-shared, thus does not leak intermediate information. Wu\;\etal\,\cite{wu20privacy} port XGBoost to the fully encrypted setting for more than two parties by using a threshold version of Paillier HE and the SPDZ framework\;\cite{damgaard2012multiparty}. Similarly to the SMC approaches, this work requires the generation of secret shares of random multiplications (Beaver triplets) by a trusted third party.
Recently, Liu\;\etal\,\cite{liu2020towards} proposed a new offloading solution that could be extended to collaborative scenarios. Using additive HE, data owners offload their data to a cloud. Using additive secret-sharing and with the help of a computing server, the cloud builds the tree from the encrypted data via tailored secure-counting and comparison protocols. They improve similar works\;\cite{li2019outsourced,li2017outsourcing}, by relaxing the need for parties to be online during the learning.

\descr{Encryption as Obfuscation.} Similar to perturbation techniques, obfuscation of the sensitive data can be achieved with encryption techniques (\eg AES)\;\cite{rao2014cryptographic,sharma2016secure,Tandel2016PrivacyPD}. The encryption can be lifted by the miner once data has been merged\;\cite{Tandel2016PrivacyPD}. Alternatively, parties offload deterministic encryption of their data, and the learning is done on the ciphertexts as new labels and feature values\;\cite{rao2014cryptographic,sharma2016secure}. But this encryption technique is prone to frequency-analysis attacks\;\cite{bellare2015secure,naveed2015inference}.

\descr{Locality-Sensitive Hashing (LSH).} SimFL\;\cite{li2019practical} includes a pre-processing phase where LSH is applied and similar information across parties is grouped without revealing raw data. LSH ensures that similar (resp. dissimilar) instances have equal (resp. different) hash digests with high probability. During training, the gradients of all similar instances are included in the boosting. SimFL improves upon previous work\;\cite{zhao2018inprivate,li2019practical} either in terms of accuracy or efficiency. While SimFL is fully distributed, it considers a relaxed threat model where a dishonest party might learn some information about the other parties through inference attacks, but not through their raw data.

\subsection{Hardware-Based Solutions}\label{sec:ppm:hw}
Trusted hardware, \eg\:secure enclaves\;\cite{mckeen2013innovative,kaplan2016amd}, is an alternative solution for private distributed model induction. In particular, a few recent works consider that each party installs a secure enclave at its premises; this enclave is responsible for storing and computing on the sensitive data\;\cite{leung2019towards,law2020secure}. Hardware-based solutions impose different trust assumptions and are orthogonal to the aforementioned software-based solutions. Also, recent research shows that secure enclaves are susceptible to side-channel attacks\;\cite{van2018foreshadow,wang2017leaky}.

\subsection{Hybrid Solutions}\label{sec:ppm:hybrid}
Hybrid solutions combine the various PETs described earlier. For example, Teng\;\etal\,\cite{teng2007hybrid} combine randomization techniques with SMC: Each party's local dataset is enhanced with perturbed data from others to find the best set of features during tree induction. This leads to a model accuracy better than randomization approaches at reduced computation costs, compared to SMC. Truex\;\etal\,\cite{truex2019hybrid} propose a hybrid approach for federated learning; it employs DP mechanisms and threshold HE. They apply it to decision-tree learning by using the ID3 algorithm and an aggregator that initiates the root node. They employ threshold additive HE on noisy inputs such that the aggregator decrypts feature counts and class counts values with DP guarantees. Their solution yields a more accurate model than others that employ local DP, as the amount of noise is divided by the number of parties required to ``unlock'' the HE threshold. Subsequently, Liu\;\etal\,\cite{liu2019boosting} improved this approach by ensuring that the aggregation is performed correctly. Moreover, combining local DP and secure aggregation with threshold HE\;\cite{bonawitz2017practical} is envisioned in both horizontal and vertical data partitioning settings for XGBoost\;\cite{tian2020federboost}.
Finally, Wu\;\etal\,\cite{wu20privacy} inject noisy values during training to achieve the guarantees of DP for the output of their SMC-based aggregation system, \ie the resulting model.

\section{Threat Model}\label{sec:tm}

We now systematize existing works on privacy-preserving collaborative decision-tree learning based on their threat model (see Table\,\ref{tab:TM}). 
We identify two main threat model categories: (a) honest and semi-honest, and (b) malicious.

\descr{Honest and Semi-Honest.} An entity is deemed honest if it abides by the protocol and does not try to infer anything from the data exchanged and stored during the protocol. A system operates under an honest model if all the parties involved are honest. While none of the surveyed works considers a fully honest threat model, many solutions rely on at least one honest entity in the system. For example, miners in the central collaborative model are de facto trusted to train the model and to inject the necessary noise to ensure differential privacy guarantees of the output, \eg\cite{friedman2001greedy,fletcher2015differentially}. 

The semi-honest model (also referred to as passive or honest-but-curious) considers that the participants follow the protocol but might try to infer as much information as possible about the other entities' private data, from the communicated values.
This is a typical threat model for solutions based on perturbation and randomization techniques that implicitly consider that parties exchange their data with a semi-honest miner. Indeed, the miner is not trusted with the original datasets but obtains full access to each party's surrogate data, \eg\cite{agrawal2000privacy, fong2010privacy}. For works that consider the application of DP in the aggregator collaborative model, the aggregator is often considered semi-honest, hence parties add noise to their local computations before sharing them with the aggregator\;\cite{liu2020federatedExtraTree,xiang2018collaborative}.

Most works using cryptographic techniques consider the semi-honest model for the different entities\;\cite{cheng2019secureboost,fang2020hybrid,gambs2007privacy,kuijpers2008privacy,lindell2000privacy,du2002building,liu2020towards,vaidya2005privacy,xiao2005privacy,xiao2006privacy}. Only a handful of these works also consider passive collusions among the different members. For instance, by using SMC frameworks or LHE, several works are secure against a collusion between half of the $N$ involved parties\;\cite{Hoogh2014practical,vaidya2008privacy}. With threshold encryption or secret sharing other works tolerate up to $N{-}2$\;\cite{gambs2007privacy,han2007multi}, or even $N{-}1$\;\cite{emekci2007privacy,vaidya2013random}, colluding participants. In both works of Truex\;\etal\,\cite{truex2019hybrid} and Wang\;\etal\,\cite{wang2020cloud}, the maximum number of colluding members admissible without damaging privacy is piloted by the threshold defined for the secret sharing scheme. In other works, however, collusion causes a direct loss of privacy: In the work of Du\;\etal\;\cite{du2002building}, the collusion between one party and the third-party assisting with the computation can reveal the other party's secret data. The revocation mechanism in the work by Liu\;\etal is secure, as long as the revoked member does not collude with the aggregator\;\cite{liu2019revocable}. Li\;\etal\,\cite{li2019practical} consider that a dishonest party might learn some information about the data of other parties, \eg local gradients, but not raw data. Theirs is one of the few works that takes into account potential leakage induced by sharing intermediate values during the training process. 

\descrit{Remark.} Note that in several works, the privacy is guaranteed by the presence of implicitly trusted third-parties involved in the generation of cryptographic keys\;\cite{li2017privacy,li2019outsourced} and random shares used for the computations (\eg Beaver triplets for multiplication operations in SMC)\;\cite{wu20privacy}.
Similarly, hardware-based solutions presume the chip manufacturer is trusted and the attacker does not have access or control over the enclave\;\cite{law2020secure,leung2019towards}.

\descr{Malicious.} Also known as active, malicious participants can \emph{actively} cheat and tamper with the protocol by crafting messages with fabricated inputs and by aiming to gain more information about the other entities' data or to simply disrupt the protocol.
A limited amount of works consider resistance against malicious entities. As a party can always tamper with its local training data to perturb the learning, the envisioned malicious model concerns active adversaries who aim to cheat the learning process by performing wrong computations. 
Emekçi\;\etal\,\cite{emekci2007privacy} rely on Shamir secret-sharing, along with a technique that verifies the correctness of an aggregate result; by including redundancy to the secret-sharing polynomials, they are able to over-determine the equation system that reveals the result upon solution. They show that with appropriate tuning, it is computationally hard for an adversary to forge a result undetected. Akavia\;\etal\,\cite{akavia2019privacy} provide privacy guarantees against a malicious aggregator that tries to learn as much as possible from the clients' inputs following any arbitrary attack strategy.
Furthermore, a malicious aggregator might be tempted to skip the correct aggregation of intermediate values to obtain some local information. This behaviour is tackled by Liu\;\etal\,\cite{liu2019boosting} with $\tau$-threshold secret-sharing which ensures that the aggregator only learns the aggregate result over the data of at least $\tau$ parties. In InPrivate\;\cite{zhao2018inprivate}, the malicious parties seek to tamper with the steepest-descent returned. Hence, the design employs a local quality control of other members' trees: Each party evaluates on its local data the performance of the tree received from the previous member and decides whether to discard it or not. 
In the work by Wu\;\etal\;\cite{wu20privacy}, malicious members that deviate from the protocol are considered as an extension of their design: They use zero-knowledge proofs and commitments to prove statements about secret data without disclosing it. In particular, each member of the collective proves that it executed the specified protocol correctly. Similarly, though not considering the malicious model, several works claim that this model can be supported using general techniques such as those presented by Goldreich\;\cite{goldreich_2004,goldreich2009foundations} at the cost of efficiency (in terms of computations and communications)\;\cite{lindell2000privacy,vaidya2013random}.
Recently, Abspoel\;\etal\,\cite{abspoel2020secure} use SPDZ\;\cite{mp-spdz} to provide a MPC-based C4.5 algorithm. Active security is achieved assuming honest majority among three non-colluding servers and sacrificing performance.

\begin{table}
\caption{\small{Threat models considered in the literature. {\protect\pie{0}}: Honest, {\protect\pie{180}}: Honest-but-curious, {\protect\pie{360}}: Malicious, Coll. Res.: Collusion resistant, $N$ number of parties, $\tau$: Secret sharing threshold, $u$: number of parties knowing the class labels, TH: Trusted Hardware.}}\vspace{-0.2cm}
\label{tab:TM}
\fontsize{8}{7.2}\selectfont
\centering
\begin{tabularx}{0.49\textwidth}{m{4.7cm}>{\centering\arraybackslash\columncolor[gray]{0.9}}cc>{\columncolor[gray]{0.9}}cc}
\toprule
Reference & \rot{\scriptsize{Parties}} & \rot{\scriptsize{Miner}} & \rot{\scriptsize{Aggregator}} & \rot{\scriptsize{Coll. Res.}}\\
\midrule
\cite{liu2020towards,li2017privacy,li2017outsourcing,ma2019privacy}& \pie{180} & \pie{180} & \pie{180}& \\ \hdashline
\cite{agrawal2000privacy,aslett2015encrypted,baghel2013privacy,brickell2009privacy,bu2007preservation,dowd2006privacy,estivill1999data,fang2010preserving,fang2009new,fong2010privacy,kadampur2010noise,kalyani2017decision,liu2009privacy,mohammed2011differentially,rao2014cryptographic,Tandel2016PrivacyPD,vadivu2014improved}& \pie{180} & \pie{180} && \\ \hdashline
\cite{akavia2019privacy} & \pie{180} & \pie{360} && \\ \hdashline
\cite{tian2020federboost,truex2019hybrid,wang2020cloud}& \pie{180} & & \pie{180}& $N{-}\tau$ \\ \hdashline
\cite{alabdulkarim2019ppsdt,alabdulkarim2019privacy,fang2008privacy,giacomelli2019privacy,liu2020federatedExtraTree,li2020multicenter,liu2019revocable,li2017outsourcedSensor,li2019outsourced,sharma2016secure,shen2009privacy,wang2006classification,xiang2018collaborative,yang2019tradeoff}& \pie{180} & & \pie{180}& \\ \hdashline
\cite{liu2019boosting} & \pie{180} & & \pie{360}& \\ \hdashline
\cite{Hoogh2014practical}& \pie{180} & && $\lceil N/2 \rceil$ \\ \hdashline
\cite{emekci2007privacy,wu20privacy,zhao2018inprivate} & \pie{360} & && $N{-}1$ \\ \hdashline
\cite{gambs2007privacy,han2007multi,sheela2013novel} & \pie{180} & && $N{-}2$ \\ \hdashline
\cite{cheng2019secureboost}& \pie{180} & && $N{-}u$ \\ \hdashline
\cite{vaidya2008privacy,vaidya2013random}& \pie{180} & && $N{-}\tau$ \\ \hdashline
\cite{behera2011privacy,dansana2013novel,du2002building,du2003using,fang2020hybrid,feng2019securegbm,gangrade2009building,gangrade2011novel,khodaparast2018privacy,kikuchi2013privacy,kuijpers2008privacy,lindell2000privacy,li2019practical,lory2012enhancing,ma2008secure,samet2008privacy,sumalatha2016fuzzy,suthampan2005privacy,teng2007hybrid,vaidya2005privacy,xiao2005privacy,xiao2006privacy,zhan2005privacy,zhan2007using} & \pie{180} & && \\ \hdashline
\cite{abspoel2020secure} & \pie{360} & && $\lceil N/2 \rceil$ \\ \hdashline
\cite{law2020secure,leung2019towards}& TH & & TH &\\ \hdashline
\cite{bai2017embedding,blum2005practical,bojarski2014differentially,consul2020differentially,fletcher2015a,fletcher2015differentially,fletcher2017differentially,friedman2010data,guan2020differentially,hou2019dprf,jagannathan2009practical,li2020privacy,liu2018differentially,nock2020boosted,patil2014differential,rana2015differentially,wang2020scalable,wu2020differentially,xin2019differentially,zhang2017dpets,zhu2013effective}& \pie{180} & \pie{0}&& \\
\midrule
\end{tabularx}
\vspace{-0.4cm}
\end{table}

\section{Leakage Analysis}\label{sec:leak}
Our analysis of the literature on privacy-preserving collaborative tree-model induction (\S\ref{sec:ppm}) shows that very few works protect the training process end-to-end. We find that most works ensure the confidentiality of the raw training data, but do not consider the leakage that might occur from computations required for collaborative tree induction, \eg comparison operations are performed on cleartext data to abide with HE limitations. To this end, we design a framework that analyzes which information is leaked during privacy-preserving collaborative tree-based model induction, enabling us to systematize the literature on that aspect.
Although several works acknowledge this leakage and even provide an analysis of their solution\;\cite{cheng2019secureboost,fang2020hybrid,gambs2007privacy,giacomelli2019privacy,liu2019boosting,vaidya2008privacy,vaidya2013random,wu20privacy}, they do not do so in a systematic way. Our framework is a first step towards a generic systematization of works based on the information leakage. We explain the need for leakage analysis and then present our taxonomy.

\subsection{Importance of Leakage Minimization}
Overall, revealing computation results during tree-based model induction, \eg statistics or model updates, produces a potential leakage about the training data. For instance, global statistics refer to values aggregated from several parties involved in the learning. This can include feature counts, class counts, the global feature list, or the party owning the best split. Wu\;\etal\,\cite{wu20privacy} describe an attack among colluding parties in the vertical data-partitioning setting that we extend here. Consider that only one party holds the class attribute. Then, colluding parties responsible for successive splits from the root can, with access to the label of the leaf, infer the class of a subset of the training data with high probability. They also propose a second attack that relies on the same principle to yield the feature value. 
Generally, during tree-based induction in the horizontal data-partitioning setting, aggregate statistics leak information about the global training set (\eg number of samples, distribution per feature). Similarly, in the vertical case, the chosen metric value for the best local feature is revealed. Even though no concrete attacks that exploit these values have been proposed, this leakage might be non-negligible and should be taken into account by designers of privacy-preserving solutions. Indeed, numerous works have shown that aggregate statistics leak information about individual samples in several settings such as genomics\;\cite{homer2008resolving}, smart metering\;\cite{buescher2017two}, and location data\;\cite{pyrgelis2017does}.

Similarly, weight updates might leak information and lead to membership or property-inference attacks\;\cite{melis2019exploiting,nasr2019comprehensive,zhu2019deep}. A curious participant could infer the presence of specific samples in other parties' data or properties about the data. This is particularly relevant for works using gradient boosting. Access to local model updates makes the training process vulnerable to updates-leak attacks\;\cite{salem2020updates} that aim at inferring information about the training sets from the model updates. Finally, as shown by Hitaj\;\etal\,\cite{hitaj2017deep}, a Generative Adversarial Network can be trained on the global aggregate updates to generate prototypical examples of the targeted victim's local dataset. Even though these attacks focus on different learning algorithms, \eg deep convolutional neural networks, these attacks show that the information communicated during gradient descent contains private data. Fang\;\etal acknowledge and tackle this leakage by using HE and secret sharing to protect the gradients end-to-end\;\cite{fang2020hybrid}.

\subsection{Leakage Taxonomy}
We now present our leakage classification of the literature on privacy-preserving collaborative tree-based models. We use the term \emph{beneficiary} for the entity in the collective who has access to the leak, \ie who is able to infer some information. This can be the miner, the aggregator, the parties themselves or a combination of these entities. We remark that training tree-based models in the collaborative setting requires communication between the different members, thus increasing potential leakage compared to a non-collaborative scenario. Three types of leakage can occur:

\descr{Data Leakage.} This directly endangers each party's raw training data. Data leakage can be the result of an improperly configured protection mechanism. For instance, works on random perturbation\;\cite{agrawal2000privacy,estivill1999data} offer no formal privacy guarantees and are vulnerable to reconstruction attacks\;\cite{kargupta2003privacy}. As mentioned in \S\ref{sec:ppm:perturb}, these approaches employ ad-hoc methods to obfuscate the training data. 
Similarly, encryption as obfuscation using deterministic encryption can provide only limited guarantees, as the relationship between data points is preserved. Works using DP mechanisms in the central model also inherently leak raw training data to the server; by design the server aggregates the raw data before applying the protection. Data leakage might also occur from the usage of a trained model. Shokri\;\etal propose a membership inference attack revealing if a sample was in the training set by looking at the classifier's prediction outputs\;\cite{shokri2017membership}. 
Similarly, inversion attacks reveal if a sensitive attribute is used in the decision-tree\;\cite{fredrikson2015model} and property-inference attacks, if the training set satisfies some properties\;\cite{ateniese2015hacking}.
We consider these attacks out of scope in this analysis; they concern the inference phase and not the training.
 
\descr{Model Leakage.} 
Tree-based models can be sensitive due to their business value and because they contain information about the data on which they were trained. For instance, Zhu and Du\;\cite{zhu2010understanding} quantify the privacy risks associated with publishing decision trees and show that a maximum entropy estimate can leak information about the training data. As a result, it is crucial for privacy-preserving solutions to protect the trained model. In our literature analysis, we identify works that protect the final model and its weights from any party, enabling only oblivious predictions on it. This is the case for a handful of works that use HE\;\cite{abspoel2020secure,aslett2015encrypted,giacomelli2019privacy,liu2019revocable,ma2019privacy,vaidya2013random}, perturbation\;\cite{kadampur2010noise}, and others that protect the model with DP mechanisms (see \S\ref{sec:ppm:DP}). Works publishing decision trees with DP guarantees introduce a different trade-off compared to cryptographic approaches. The privacy budget determines the amount of admissible leakage for the model and training data, at the cost of accuracy.

\descr{Leakage of Intermediate Values.}
As discussed in \S\ref{sec:ppm}, numerous works tolerate the exchange of cleartext information during training to cope with challenging cryptographic operations, \eg comparisons and gains. We identify two classes depending on whether the revealed information is local (\ie related to only one party) or global (\ie aggregated over multiple parties). The type of leaked intermediate information varies depending on the learning algorithm. For standard-tree learning algorithms such as ID3, this information represents the selected split (\ie feature and split value) but also the gains and counts per tree path. For completely random approaches, the intermediate values are the class counts at the leaves. Other statistics, such as the number of local samples reaching a specific node, residual accuracy errors\;\cite{feng2019securegbm}, the ``owner'' of a node in the vertically distributed scenario (\ie which party holds the selected attribute for the node), might also be leaked. For gradient boosting ensemble models, the intermediate values are the weight updates or residuals. As shown in recent work by Hitaj\;\etal for deep models, the use of a DP mechanism can have a very limited effect on the privacy when such intermediate values are shared\;\cite{hitaj2017deep}. 

\descr{Summary.} Following our observations, we systematize the different works in the literature. For each work, we first identify the privacy-preserving mechanisms employed. Then, we analyze the type of data exchanged during the learning process and under which form (\eg encrypted, blinded, aggregated, noisy, or in clear). We assess if this corresponds to leakage of (local or global) intermediate values that we classify as statistics or weights depending on the type of data and the learning algorithm. Finally, we identify which entity gains more information from the leak (\ie the beneficiary of the leak). We introduce an additional category for works that reveal the local model to the aggregator\;\cite{alabdulkarim2019privacy,li2020multicenter,xiang2018collaborative}.
Finally, we describe how each work suffers from the leakage of information. Our analysis identifies three major groups: (a) works that do not leak any intermediate values, (b) works that leak intermediate values with local DP guarantees, and (c) works that leak intermediate statistics or weights.
We present our findings in Table\;\ref{tab:leak}. 
For conciseness, we do not display works envisioned in the central collaborative model that use DP mechanisms or perturbation techniques for privacy preservation, as we have already analyzed their leakage above in the data leakage description.
Overall, we find that only a few cryptographic works keep the final model secret\;\cite{akavia2019privacy,aslett2015encrypted,Hoogh2014practical,giacomelli2019privacy,leung2019towards,liu2019revocable,ma2019privacy,vaidya2013random}. 
Additionally, our framework shows that, though most works of Table\;\ref{tab:leak} rely on the same PETs (\eg secret sharing or HE), only a few of them avoid leaking intermediate values\;\cite{abspoel2020secure,aslett2015encrypted,li2017privacy,lindell2000privacy,lory2012enhancing,ma2019privacy,vaidya2013random,wu20privacy,xiao2005privacy,xiao2006privacy}. This is probably due to the fact that ensuring \textit{zero-leakage} during tree-based model induction comes with high communication and computation overheads.

\begin{table*}[!htb]
\caption{Intermediate value leakage in the literature. DP: Differential Privacy, P: Party, A: Aggregator, M: Miner.}\vspace{-0.2cm}
\label{tab:leak}
\fontsize{8}{7.2}\selectfont
\centering
\begin{tabular}{l|l|>{\centering\arraybackslash}p{0.4cm}>{\columncolor[gray]{0.9}\centering\arraybackslash}p{0.4cm}>{\centering\arraybackslash}p{0.4cm}|>{\columncolor[gray]{0.9}\centering\arraybackslash}p{0.8cm}>{\centering\arraybackslash}p{0.6cm}>{\columncolor[gray]{0.9}\centering\arraybackslash}p{0.6cm}>{\centering\arraybackslash}p{0.6cm}>{\columncolor[gray]{0.9}\centering\arraybackslash}p{0.6cm}>{\centering\arraybackslash}p{0.6cm}|l}
 \toprule
 \multicolumn{1}{l}{}&\multicolumn{1}{l}{} &\multicolumn{1}{l}{}&\multicolumn{1}{l}{}&\multicolumn{1}{l}{}& \multicolumn{6}{c}{What is leaked}&\\[-0.2em]
 \cmidrule(lr){6-11}
 \multicolumn{1}{l}{}&\multicolumn{1}{l}{} & \multicolumn{3}{c}{Beneficiary\;}&\multicolumn{1}{c}{Perturb.}&\multicolumn{1}{c}{Loc.}& \multicolumn{2}{c}{Statistics}& \multicolumn{2}{c}{Weights} &\\[-0.2em]
 \cmidrule(lr){3-5} \cmidrule(lr){8-9} \cmidrule(lr){10-11}
\vspace{-0.4em} Ref. & \multicolumn{1}{c}{PET} & \multicolumn{1}{c}{P}& \multicolumn{1}{c}{A} & \multicolumn{1}{c}{M} & \multicolumn{1}{c}{values}& \multicolumn{1}{c}{Model}& \multicolumn{1}{c}{Loc.}& \multicolumn{1}{c}{Glob.} & \multicolumn{1}{c}{Loc.}& \multicolumn{1}{c}{Glob.} & Description\\[0.2em]
\midrule
\cite{li2019practical}& LSH  & \pie{360} &&&&& \pie{360} && \pie{360} && Table of similar instances and local gradients \\
\cite{dansana2013novel} & LHE & \pie{360} &&&&& \pie{360} &&&& Best local gain and node owner \\
\cite{feng2019securegbm}& HE+OT & \pie{360} &&&&& \pie{360} &&&& $P_1$ learns residual errors from $P_2$\\
\cite{gambs2007privacy} & SMC\cite{kissner2005privacy} & \pie{360} &&&&\pie{360}& &\pie{360}&&& Owner, number, or local best weak classifier \\
\cite{gangrade2009building} & LHE & \pie{360} &&&&& \pie{360} &&&& Best local gain\\
\cite{teng2007hybrid} & random+SMC & \pie{360} &&&&& \pie{360} &&&& Best k attributes on augmented dataset to P\\
\cite{akavia2019privacy}& FHE & \pie{360} &&&&&& \pie{360} &&& All gains\\
\cite{alabdulkarim2019ppsdt}& LHE+Blindings & \pie{360} &&&&&& \pie{360} &&& All gains\\
\cite{behera2011privacy}& SMC & \pie{360} &&&&&& \pie{360} &&& All gains\\
\cite{du2002building} & SMC+SS & \pie{360} &&&&&& \pie{360} &&& Depends how max computed \\
\cite{emekci2007privacy}& SSS & \pie{360} &&&&&& \pie{360} &&& Conditional entropy, global class count\\
\cite{khodaparast2018privacy} & LHE & \pie{360} &&&&&& \pie{360} &&& Structure and counts of every class per leaf\\
\cite{kikuchi2013privacy} & LHE+SS & \pie{360} &&&&&& \pie{360} &&& Who owns best attribute\\
\cite{kuijpers2008privacy}& SMC & \pie{360} &&&&&& \pie{360} &&& Tree structure, best gain and node owner\\
\cite{li2017outsourcing}& HE & \pie{360} &&&&&& \pie{360} &&& Best attribute \\
\cite{li2019outsourced} & LHE (thrs) & \pie{360} &&&&&& \pie{360} &&& All or best gain\\
\cite{ma2008secure} & SS+GC & \pie{360} &&&&&& \pie{360} &&& Sign of secret\;\\
\cite{samet2008privacy} & LHE+SMC & \pie{360} &&&&&& \pie{360} &&& Global square ratio\\
\cite{sheela2013novel}& SMC+SS & \pie{360} &&&&&& \pie{360} &&& Best local gain and node owner \\
\cite{wang2020cloud}& SS && \pie{360} &&&& \pie{360} & \pie{360} && \pie{360} & Global gradients ($h_i$,$g_i$)\\
\cite{liu2019revocable} & LHE (thrs) && \pie{360} &&&& \pie{360} &&&& Local Split vector W to A\\
\cite{vaidya2005privacy}& LHE && \pie{360} &&&& \pie{360} &&&& Class distribution of each node, local best gain\\
\cite{zhan2005privacy}& LHE+Blindings && \pie{360} &&&& \pie{360} &&&& Leader learns permuted decryption\\
\cite{zhan2007using}& LHE+Blindings && \pie{360} &&&& \pie{360} &&&& Leader learns permuted decryption\\
\cite{cheng2019secureboost} & LHE && \pie{360} &&&&& \pie{360} && \pie{360} & Residuals, node owner, split, and $d$ \\
\cite{fang2008privacy}& SMC && \pie{360} &&&&& \pie{360} &&& Node owner and index of best feature \\
\cite{gangrade2011novel}& SMC && \pie{360} &&&&& \pie{360} &&& Global gain for each feature to Leader \\
\cite{han2007multi} & LHE && \pie{360} &&&&& \pie{360} &&& Gain value for all features to Leader\\
\cite{liu2020towards} & LHE+SS (thrs) && \pie{360} &&&&& \pie{360} &&& Gain value for all features\\
\cite{shen2009privacy}& SMC && \pie{360} &&&&& \pie{360} &&& Gain value for all features\\
\cite{sumalatha2016fuzzy} & LHE && \pie{360} &&&&& \pie{360} &&& Global class distribution to Leader\\
\cite{truex2019hybrid}& DP+HE && \pie{360} &&&&& \pie{360} &&& Global gradients ($h_i$,$g_i$)\\
\cite{vaidya2008privacy}& LHE && \pie{360} &&&&& \pie{360} &&& Best site, local best gain \\
\cite{liu2019boosting}& LHE+SS (thrs) && \pie{360} &&&&&&& \pie{360} & Global gradients ($h_i$,$g_i$) \\
\cite{alabdulkarim2019privacy}& Electronic Voting && \pie{360} &&& \pie{360} &&&&& Local ensemble \\
\cite{li2020multicenter}& DP-GAN && \pie{360} &&& \pie{360} &&&&& locally trained optimal tree \\
\cite{fang2009new}& LHE &&& \pie{360} &&&& \pie{360} &&& All gains\\
\cite{fang2010preserving} & LHE &&& \pie{360} &&&& \pie{360} &&& All gains, node owner \\
\cite{zhao2018inprivate}& DP mechanism& \pie{360} &&&  &\pie{360}&&&&& Local DP models \\
\cite{liu2020federatedExtraTree}& DP mechanism&& \pie{360} && \pie{360} &&&&&& DP protected intermediates \\
\cite{xiang2018collaborative} & DP mechanism&& \pie{360} &&& \pie{360} &&&&& Local DP models\\
\cite{tian2020federboost} & DP+LHE (thrs) & & \pie{360} && \pie{360} &&&&&& Perturbed global gradients (more with collusion)\\
\bottomrule
\end{tabular}
\vspace{-0.3cm}
\end{table*}

\section{Evaluation and Comparison}\label{sec:eval}

Overall, fair comparisons between works are non-trivial, as the learning, collaborative, and privacy models are orthogonal, and the choice of evaluation metrics and datasets might differ. Additionally, very few works provide a public implementation which makes comparisons harder. Several datasets are commonly used for benchmarking the solutions (\eg \cite{UCI,credit1,credit2}), but unfortunately, they often are too limited in the number of features and samples to be realistic for modern big-data tasks (see Appendix~\ref{app:eval} for more details).

A formal comparison of communication and computation overhead in the literature is challenging to achieve as works seldomly vary only one of the four axes we have identified. Nonetheless, several conclusions can be drawn. 
The communication overhead is constrained by the collaborative model, the type of data partitioning, and the number of nodes in the tree: For instance, in the fully distributed models with $N$ parties and horizontal data partitioning, local information needs to be exchanged with all other parties, whereas in the aggregator model, only the aggregator needs to be informed, yielding a communication in $O(N^2)$ versus $O(N)$, respectively (c.f. \cite{emekci2007privacy} vs. \cite{aslett2015encrypted}). Conversely, in the vertical setting, communication is less affected, as the difficulty lies only in finding the best party to split the data (which can be achieved by only sharing the best locally computed gain and not intermediate values such as counts).  
With respect to computation overhead, the horizontal data-partitioning setting requires each party to compute local intermediates: Instead of computing this information once, it is computed locally by each party on smaller datasets. Eventually, all the intermediates are exchanged and aggregated to find the best split.
Protection mechanisms also play a crucial role in the overhead. 
Works employing differentially private mechanisms, randomization or perturbation techniques report minimal to no computation and communication overhead as the distributed learning process operates similarly as if the data was not perturbed. Hardware-based solutions present a computational overhead ranging between $4.5\times$ and $5.1\times$ to perform encrypted XGBoost compared to the standard algorithm\,\cite{law2020secure}.
Cryptographic solutions have an even more dramatic impact on computational and communication overhead. 
For fully distributed tree induction with horizontal partitioning, SMC solutions induce a communication overhead of $O(N^2)$ (\eg \cite{Hoogh2014practical,kuijpers2008privacy}) and while some HE solutions have a similar asymptotic overhead (\eg \cite{samet2008privacy,xiao2005privacy}), others can reduce this overhead to $O(c^2+Nc)$ (\eg \cite{zhan2005privacy}), where $c$ is the number of class labels. However, the more complex homomorphic operations imply a bigger computational overhead, \eg an overhead of $O(N^2|A|)$ \cite{samet2008privacy,xiao2005privacy} vs. $O(|A|)$ for a pure SMC solution \cite{emekci2007privacy}, with $|A|$ the number of attributes depending only on the training set.
Recently, Wu \etal\;\cite{wu20privacy} showed that combining HE and SMC can reduce the computational overhead achieving up to a $19.8\times$ speed up compared to a pure SMC solution. Indeed, in their approach, the secret-shared values among the parties are reduced from $O(n|A|)$ to $O(b|A|)$ with $b$ the number of split values for any feature. 

\section{Open Challenges}\label{sec:open}
Our deep analysis of privacy-preserving collaborative solutions for decision-tree induction enables us to identify some open problems and challenges in this field.

\descr{Motivating Privacy-Preserving Induction and Quantification of Leakage.}
As discussed in \S\ref{sec:tm}, the motivation for privacy-protection in collaborative tree induction is not clearly stated in the literature; there exist no papers that investigate privacy attacks that can occur while training a tree-based model in a federated setting. On the contrary, research on other learning algorithms, \eg neural networks, demonstrates several attacks, \eg membership or property-inference ones\;\cite{melis2019exploiting,nasr2019comprehensive,hitaj2017deep}, that might take place in collaborative settings. Indeed, the need for strong protection of intermediate computation values during tree induction has not been extensively studied. Hence, the design of more efficient protocols for privacy protection requires an understanding of the quantitative effect of such leakage on the data and model privacy. Our leakage analysis framework is a first step towards this research direction. 

\descr{End-to-End Protection.}
Although numerous cryptographic solutions on privacy-preserving distributed decision-tree learning exist, very few of them provide full privacy protection by safeguarding the parties' data, the training process, its outputs, and preventing intermediate value leakage. Works that achieve this protection level cover only some collaborative models and learning algorithms. For instance, random forests and CRTs have only been covered in the offloading\;\cite{aslett2015encrypted,ma2019privacy} and fully-distributed\;\cite{vaidya2013random} scenarios. The end-to-end solution by De Hoogh\;\etal works only for categorical data\;\cite{Hoogh2014practical} and the work lifting this limitation operates in the offloading model\;\cite{abspoel2020secure}. 
Thus, hybrid solutions combining HE or secret-sharing with differential privacy to trade-off some accuracy for end-to-end privacy protection in collaborative tree-based learning is a promising direction\;\cite{wu20privacy}.

\descr{Malicious Members in the Collective.}
In \S\ref{sec:tm}, we analyzed the literature's threat models and found that most works consider semi-honest entities. Whereas this model is deemed acceptable for research prototypes, it might not reflect real-world collaborative scenarios involving multiple entities; some of them might have interest in exploiting the collaborative training process for their own benefits. To this end, it would be valuable to design solutions in stronger models, \eg the anytrust model\;\cite{wolinsky2012scalable} or a fully malicious one. While some works have addressed the former, they only consider limited collaborative models and learning algorithms, \eg \cite{emekci2007privacy,wu20privacy}.
Addressing fully malicious models could be achieved by employing additional cryptographic techniques, \eg verifiable computation and zero-knowledge proofs, or robust learning techniques that prevent malicious participants from actively biasing the learning. 
For instance, zero-knowledge proofs can ensure the correct execution of decision-tree predictions on a secret model held by one party\;\cite{zhang2020zero}. Fiore\;\etal\,\cite{fiore2014efficiently, fiore2020boosting} apply similar techniques for ensuring correct computations on outsourced encrypted data. Alternatively, works in the direction of \emph{Byzantine tolerance} propose statistical techniques that prevent malicious parties from poisoning the learning process\;\cite{elmhamdi2020genuinely}; these techniques could be combined with cryptographic protocols to get the best of both worlds. As noted by Bagdasaryan\;\etal\,\cite{bagdasaryan2020backdoor}, this is an open problem for the PETs community not only for tree-based models but for federated learning in general. We remark, however, that decision trees have a substantially different structure than neural networks and that no work has ever considered byzantine attacks for collaborative tree-based models. Consequently, understanding how byzantine parties can disrupt the learning of distributed tree-based induction is a crucial first step.

\descr{Resilience to Dropout and Fault Tolerance.}
Party dropouts occur when a member of the collective decides to leave the consortium while the training process has not been finalized. Following data protection laws, such as the GDPR\;\cite{gdpr}, it is desirable that this member's data is not taken into account for the model training. This functionality, also referred to as the \emph{right to be forgotten}, can jeopardize privacy, in particular for tree-based models\;\cite{chen2020machine}. 
Yet, not many works address this problem\;\cite{liu2019boosting,wang2020cloud}. A possible deterrent is the fact that dropouts are incompatible with widely used cryptographic solutions such as HE or secret sharing where all parties are needed to obtain a computation result. Threshold cryptosystems could offer a solution to the problem as the result of the computation can be obtained with the participation of only a subset of entities.

Fault tolerance relates to communication issues that arise when members of the collective are unavailable during the training (or respond asynchronously to queries); the unreliability of the network and potential hardware limitations of the parties could impede collaborative learning by waiting for the unavailable parties. 
For instance, the learning can be \textit{stuck} if the aggregator requires a sufficient number of local updates to proceed with its computations, if the collective relies on a leader for protocol computations\;\cite{zhan2005privacy}, or if a member is unavailable in the sequential setting\;\cite{zhao2018inprivate}. A potential research direction is again the application of threshold cryptography and its combination with learning algorithms that cope with asynchronicity.

\section{Conclusion}\label{sec:conc}
In this work, we have systematized the knowledge stemming from the literature on privacy-preserving collaborative tree-based model induction. We observed that preserving privacy in the distributed setting incurs challenges that most solutions address by modifying the learning algorithm or introducing new entities in the collective. Numerous privacy-preserving technologies have been employed to account for combinations of learning algorithms, collaborative settings, and threat models; hybrid solutions that combine cryptographic protocols with differential privacy are the most promising approach for achieving end-to-end privacy protection for collaborative tree model induction. We have introduced a leakage identification framework that enables protocol designers to categorize and reason about the leakage that stems from the intermediate information computed and exchanged in collaborative model training. Our systematization and framework highlighted several avenues for further research on the topic such as support for malicious models, asynchronicity, and leakage protection.

\section{Acknowledgements}
We would like to thank our shepherd Sebastian Meiser and the anonymous reviewers for their helpful feedback. We are also grateful to Giovanni Cherubin, Andrew Janowczyk, Peizhao Hu, and the members of the EPFL Laboratory for Data Security for their helpful comments and suggestions. This work was supported in part by the grant C17-16 (SecureKG) of the Swiss Data Science Center.

\Urlmuskip=0mu plus 1mu\relax
\bibliographystyle{IEEEtranS}
\bibliography{references}

\begin{appendices}

\section{Experimental Evaluations}\label{app:eval}
We now expand our analysis with respect to the evaluations performed by the surveyed works.

\descr{Model Evaluation Metrics.} Most works evaluate their solutions against the model's accuracy, error, area under the curve or F1 score on standard open source datasets\;\cite{UCI,credit1,credit2,credit3}. Most DP-based works estimate the model's accuracy as a function of the privacy budget enabling cross-work comparison\;\cite{zhang2017dpets, jagannathan2009practical,patil2014differential}. Another popular metric for cryptographic solutions is the training execution time that is evaluated against the number of trees, the depth, or the number of parties.

\descr{Datasets.} The UCI database\;\cite{UCI} and Kaggle\;\cite{credit1,credit2,credit3}, are the predominant sources of the datasets used in the literature. Although this facilitates comparisons, we note that several of these datasets contain a very limited number of features or very few samples, hence they might not be representative of modern big-data tasks performed today.

\descr{Open Source Implementations.} Finally, very few works provide a public implementation of their solution\;\cite{bojarski2014differentially,consul2020differentially,fletcher2015a,fletcher2015differentially,fletcher2017differentially,leung2019towards,yang2019tradeoff}; this makes reproduction and comparison with related works harder. 

\section{Literature Analysis}\label{app:analysis}
\begin{figure}[t]
 \centering
 \includegraphics[width=.45\textwidth]{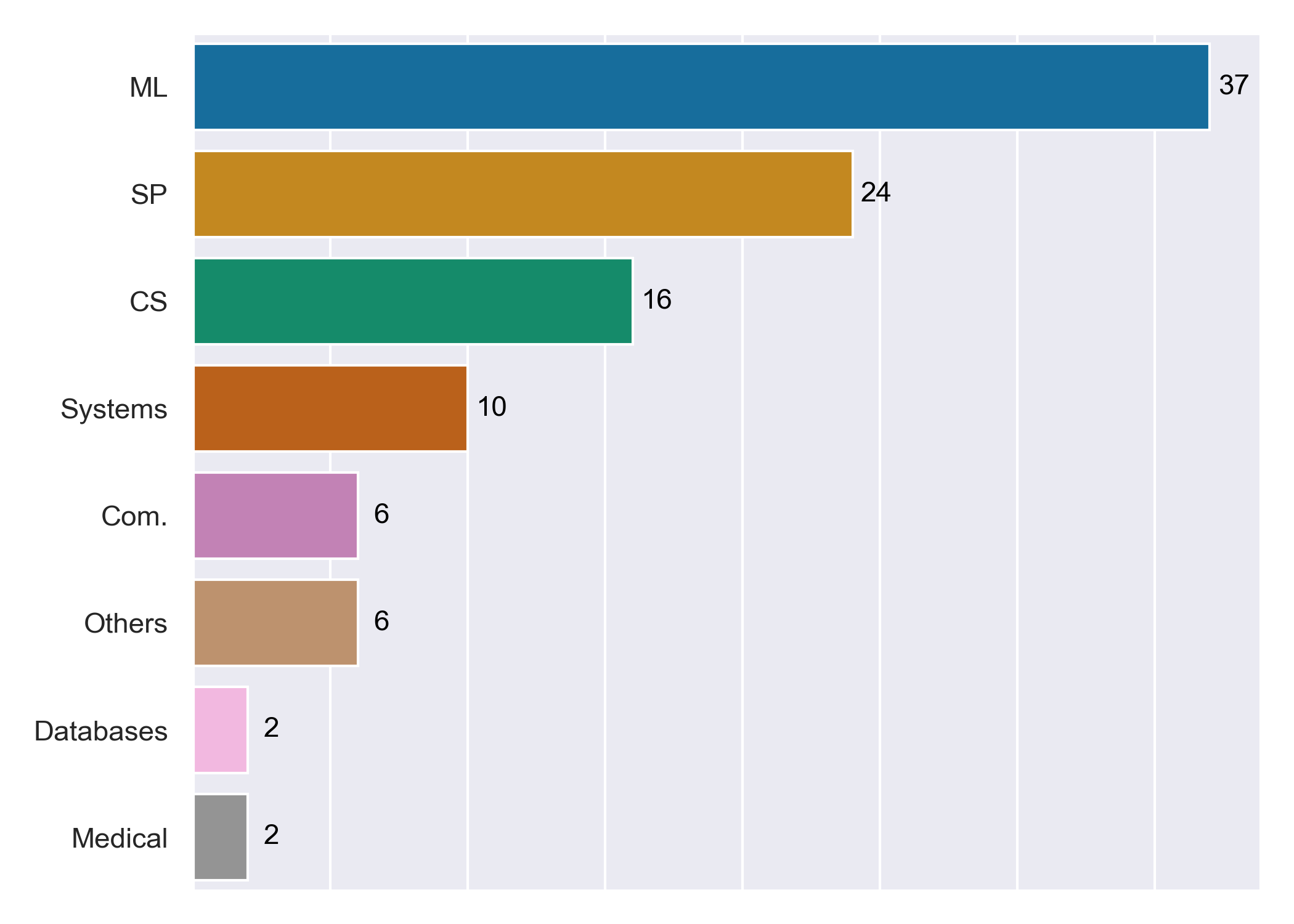}
 \caption{Number of surveyed works across the different communities.}
 \label{fig:community}
\end{figure}

In this section, we provide further details about the different works studied in this paper. As discussed in \S\ref{sec:method}, we used Google Scholar\;\cite{scholar}, Microsoft Academic\;\cite{microsoft}, and DBLP\;\cite{dblp} to discover works related to combinations of the different learning algorithms with the following key words: {\textit{\{private, privacy-preserving, collaborative, distributed, training, induction, learning\}}}. Then, we cross-referenced the different works to expand the search.
Following our search methodology we discovered 103 works originating from a wide variety of communities. Figure~\ref{fig:community} shows the distribution of the investigated works across their communities. We find that most of the works come from the \emph{machine learning}, and \emph{data mining} communities (ML in Fig.~\ref{fig:community}): These account for $36$\% of the investigated works. The \emph{security, privacy}, and \emph{cryptography} community (SP) hosts $23$\% of the works with 24 publications. General \emph{computer science} journals are also well represented with 16 works (dubbed CS in Fig.~\ref{fig:community}). \emph{Systems} cover a wide range of fields from distributed to intelligent systems and represent $10$\% of the literature on collaborative privacy-preserving tree-based model learning. The \emph{communications}, \emph{medical}, and \emph{database} communities account respectively for $6$\%, $2$\%, and $2$\% of the surveyed works. We regroup the remaining works as \emph{others} (\eg ubiquitous computing, signal processing, etc.). At the time of writing, the ML and SP communities appear as the most influential ones on the topic, yielding more than $94$\% of the citations across all communities.

\end{appendices}

\end{document}